\documentclass[twocolumn,nofootinbib,floatfix]{revtex4-1}

\usepackage{graphicx}
\usepackage{bm}
\usepackage{gensymb}
\usepackage{color}
\usepackage{soul}
\usepackage{amsmath}
\usepackage{amssymb}

\begin{document}

\title{
Structural evolution of amorphous polymeric nitrogen from \textit{ab initio} molecular dynamics simulations and evolutionary search
}

%%%%%%%%%%%%%%%%%%%%%%%%%%%%%%%%%%%%%%%%%%%%%%%%%%%%%%%%%%%%%%%%%%%%%%%%%%%%%%%%%%%%%%%%%%%%%%%%%%%%%%%%%%%%%%%%%%%%
\author{Dominika Melicherov\'{a}}
\email{dominika.melicherova@fmph.uniba.sk}
\affiliation{Department of Experimental Physics, Comenius University in Bratislava, Mlynsk\'{a} Dolina F2, 842 48 Bratislava, Slovakia}

\author{Oto Kohul\'{a}k}
\affiliation{Department of Experimental Physics, Comenius University in Bratislava, Mlynsk\'{a} Dolina F2, 842 48 Bratislava, Slovakia}

\author{Du\v{s}an Pla\v{s}ienka}
%\email{plasienka@fmph.uniba.sk}
\affiliation{Department of Experimental Physics, Comenius University in Bratislava, Mlynsk\'{a} Dolina F2, 842 48 Bratislava, Slovakia}

\author{Roman Marto\v{n}\'{a}k}
\affiliation{Department of Experimental Physics, Comenius University in Bratislava, Mlynsk\'{a} Dolina F2, 842 48 Bratislava, Slovakia}

\date{\today}

\begin{abstract}
Polymeric nitrogen with single bonds can be created from the molecular form at high pressure and due to large energy difference between triple and single bonds it is interesting as an energetic material. Its structure and properties are, however, still not well understood. We studied amorphous nitrogen by \textit{ab initio} simulations, employing molecular dynamics and evolutionary algorithms. Amorphous nitrogen was prepared at a pressure of 120 GPa by quenching from a hot liquid, by pressure-induced amorphization of a molecular crystal, and by evolutionary search. All three amorphous forms were found to be structurally similar. We studied in detail the structural evolution of the system upon decompression from 120 GPa to zero pressure at 100 K. At pressures above 100 GPa, the system consists mainly of 3-coordinated atoms (80 \%) connected by single bonds while some short chains made of 2-coordinated atoms are also present. Upon decompression, the number of 3-coordinated atoms rapidly decreases below 60 GPa and longer chains are created. At 20 GPa the system starts to create also N$_2$ molecules and the ultimate structure at $p=0$ contains molecules inside a polymeric network consisting dominantly of longer chains made of 2-coordinated atoms. Besides structure, we also study vibrational and electronic properties of the system and estimate the amount of energy that could be stored in amorphous nitrogen at ambient pressure.
\end{abstract}

%%%%%%%%%%%%%%%%%%%%%%%%%%%%%%%%%%%%%%%%%%%%%%%%%%%%%%%%%%%%%%%%%%%%%%%%%%%%%%%%%%%%%%%%%%%%%%%%%%%%%%%%%%%%%%%%%%%%
\maketitle

%%%%%%%%%%%%%%%%%%%%%%%%%%%%%%%%%%%%%% INTRODUCTION %%%%%%%%%%%%%%%%%%%%%%%%%%%%
\section{Introduction}

Nitrogen is one of the most important elements, abundant in the Earth's atmosphere as well as in the solar system. Its phase diagram is complex and contains a number of molecular phases (for a review see Ref.\cite{Katzke}). The diatomic molecule N$\equiv$N has a unique property of a large binding energy of 4.88 eV/atom due to the presence of a very strong triple bond \cite{Uddin, Fuchs}. Long ago it was suggested that this triple bond can be destabilized at high pressure where the molecular crystal would be replaced by a polymeric phase with each atom having three single bonds \cite{Martin-Needs1986}. As a possible crystal structure for such phase the cubic-gauche (cg-N) structure was proposed which represents, in a sense, an analog of the diamond structure for trivalent atoms \cite{Mailhiot1992}. The large energy difference between the single-bond polymeric cg-N and triple-bond molecular N$_2$ crystal, estimated at zero pressure to about 1.4 eV/atom \cite{Uddin}, makes polymeric nitrogen interesting as a potential energetic material. According to theoretical predictions cg-N at $T=0$ should become thermodynamically more stable than a molecular crystal at pressure above 50 GPa \cite{Mailhiot1992}. The experimental preparation of cg-N is, however, not easy, due to large kinetic barriers. The predicted phase was finally experimentally confirmed in Refs. \cite{Eremets-N-2004-1, Eremets-N-2004-2} where it was synthesized by considerable overpressurization of the molecular phase to above 110 GPa at high temperature over 2000 K. It was also shown that this phase cannot be decompressed to ambient pressure since at room temperature it converts at 42 GPa back to the molecular phase \cite{Eremets-N-2004-1}. The mechanism of creation of cg-N from the molecular phase was theoretically studied by \textit{ab initio} metadynamics in Ref.\cite{Plasienka2015}. Several recent works investigated the possibility of creating and stabilizing single N-N bonds in systems other than pure nitrogen \cite{Williams2017,Steele2017,Steele2017_2,Steele2017_3,Steele2017_4,zhang2017}.

Besides the cg-N phase, also amorphous polymeric N (a-N), called the $\eta$-phase, was observed in experiments starting from a molecular crystal upon compression to about $p=150$ GPa at room temperature \cite{Goncharov-N-2000, Eremets-N-2001, Gregoryanz-N-2001}. A possibly different amorphous form with reddish color was observed by direct laser heating of the molecular form to 1400 K at 120-130 GPa in Ref.\cite{Lipp}. Importantly, a-N at low temperature below 100 K was shown to be quenchable to ambient pressure \cite{Eremets-N-2001}. Since the property of main interest here is the stored energy, which is not related to crystallinity but rather to the number of single bonds, the amorphous version of polymeric nitrogen is certainly worth study, not only for fundamental but also practical reasons. In fact, it is not obvious that the structure of a-N represents a plain disordered analog of the cg-N phase and it might include some structural motifs from low-lying metastable phases as well. In Ref.\cite{Goncharov-N-2000} it was shown, based on optical measurements and analysis of the Urbach tail, that at $p=160$ GPa a-N has a coordination number of about 2.5 suggesting that it might consist of a mixture of 2- and 3-coordinated atoms. In Ref.\cite{Goncharov-N-2008} it was suggested that there might be a connection between the negative slope of the liquid-solid phase boundary in nitrogen (in a pressure range of 66-87 GPa) and the existence of a-N, similar to the existence of amorphous water ice \cite{Tse1999}. According to this interpretation the amorphous phase represents a product of mechanical melting of the parent crystalline phase occuring upon crossing the metastable extension of the melting line.

Polymeric a-N was studied theoretically in Refs. \cite{Mattson-PhD2003, Nordlund, Boates-Bonev-N-2011, Yakub2003, Yakub2009, Yakub2016}. In an early study \cite{Mattson-PhD2003} it was investigated by \textit{ab initio} molecular dynamics (MD) simulations employing a supercell with 64 atoms and a simulation time of about 10 ps. It was suggested that the average coordination of a-N is between 2.2 and 2.4, slightly less than in experiment \cite{Goncharov-N-2000}. In Ref.\cite{Nordlund} a simulation study was performed using a classical potential, finding that a-N at ambient pressure consists mainly of 3-coordinated atoms. This study, however, compressed the system to a pressure of 1100 GPa and the applicability of a classical force field at such extremely high pressures might not be justified. In a more recent \textit{ab initio} MD study \cite{Boates-Bonev-N-2011} a-N was prepared by cooling from liquid and a mixture of 2 and 3-coordinated atoms was found. The analysis of a-N focused mainly on the Peierls distortion of the polymeric chains and electronic DOS in the range of pressure from 90 to 330 GPa. To our knowledge, a comprehensive analysis of structural evolution upon decompression from pressure of order of 100 GPa down to ambient pressure based on state-of-the-art \textit{ab initio} simulations has not been performed for a-N. We mention that \textit{ab initio} MD studies of structural evolution of amorphous phases of sulfur and CO$_{2}$ were performed by some of us in Refs. \cite{PhysRevB.85.094112} and \cite{PhysRevB.89.134105}.

In order to fill this gap we study here the structure and properties of a-N by \textit{ab initio} MD simulations and evolutionary algorithms. The paper is organized as follows. In Sec. \ref{preparation} we describe several approaches to preparation of amorphous structures by computer simulations, focusing on the applicability of evolutionary algorithms. In Sec. \ref{simulations} we present the details of \textit{ab initio} calculations and describe the protocols employed to create amorphous nitrogen. Section \ref{results} is devoted to analysis of the properties of amorphous structures and their evolution upon decompression to ambient pressure and subsequent molecularization. In the final, Sec. \ref{conclusions}, we summarize the results and draw conclusions.

%%%%%%%%%%%%%%%%%%%%%%%%%%%%%%%%%%%%%%%%%%% RESULTS %%%%%%%%%%%%%%%%%%%%%%%%%%%%%%%%%%%%%%%%%%%%%%%%%
\section{Preparation of amorphous structures}
\label{preparation}

The preparation of amorphous structures via \textit{ab initio} simulations represents a non-trivial task for several reasons. First of all, periodic boundary conditions that are routinely used impose on the disordered system an unphysical periodicity and this effect can only be mitigated by making the simulation supercell sufficiently large. Second, the short timescale available in \textit{ab initio} simulations necessarily implies that any structural transformation simulated has to be extremely fast compared to experimental timescales. As discussed in Ref.\cite{Beaudet-Mattson-Rice}, the standard method of quenching the liquid suffers from the drawback of dependence of the final structure on the cooling rate; if the latter is too fast, the resulting amorphous structure is likely to "remember too much" of the structure of the hot liquid. Similar problem arises in the case of pressure-induced amorphization, where the amorphous structure prepared by rapid amorphization might "remember" too much of the parent crystalline structure. In principle, one might attempt also to simulate the heating of the compressed molecular form similarly to the experimental procedure in Ref.\cite{Lipp}.  In order to avoid computational artifacts it would be highly desirable to have an independent method allowing one to prepare amorphous structures "from scratch", without relying on a particular initial liquid or crystalline phase, such as ~the Monte Carlo bond-switching method\cite{CRNmodel}.

In this work we approached the problem of finding the high-pressure structure of amorphous polymeric nitrogen by three independent simulation methods.  Two of them - formation of the amorphous form by melt quenching (glass) (1) and by pressure-induced amorphization (2) represent real physical experiments and in our work were simulated by means of \textit{ab initio} MD at constant pressure.  The third one is the evolutionary structure searching method (3) in combination with \textit{ab initio} total-energy calculations that aims at finding a structure with minimal enthalpy. We prepared and analyzed three amorphous states of nitrogen at $p=120$ GPa employing the above methods and compared them to each other and to available experimental data.

The application of evolutionary algorithms (EAs) to crystal structure prediction \cite{Oganov-Glass-USPEX, XtalOpt} was shown to be highly successful in numerous cases \cite{Oganov-Glass-USPEX, wang2015, rakitin2015, dong2017, kvashnin2017, zhang2017}. For systems with not too large unit cells, counting up to 20 - 40 atoms it is now a routine task to determine the lowest energy or enthalpy structure, employing evolutionary algorithms in combination with \textit{ab initio} total energy calculations. With increasing cell size, however, finding the crystalline structure becomes a problem because of too large dimensionality of the search space and the necessity of generating an astronomical number of structures to reach the crystalline ground state. This can also be interpreted as failure of the search to reach the ground state because of excessive computational complexity of the problem. Intuitively, in such case the evolutionary algorithm is still likely to produce low-energy (or enthalpy) structures, which, however, do not exhibit a long-range crystalline order. Such structures are likely to represent good disordered or amorphous structures and it is therefore plausible to expect that an efficient evolutionary algorithm applied to a large unit cell can be employed to search for such structures. In good glass formers the system is protected from crystallization by the timescale gap between experimental cooling time and much longer time required for crystallization. Similarly, even if in EAs there is no concept of time, a sufficiently large system is protected from crystallization by an excessively large number of structures necessary to find the crystalline ground state. We note that while in the search for crystal structures the space group symmetries are commonly used we do not make use of these symmetries in our search as it would contradict our aim of finding non-crystalline amorphous phases.

To our knowledge, the only application of this approach so far is Ref.\cite{Nahas-EA} where amorphous forms of silicon and indium gallium zinc oxide were found by using relatively small supercells with 64 and 84 atoms, respectively. The generated disordered structures were compared to the ones found from \textit{ab initio} MD melt quenching and from experiment focusing on local quantities such as coordination, bond lengths, and bond angles. This reveals information about local order but not about medium-range order since, e.g.,~the bond length represents a stiff degree of freedom and it is by necessity very similar in both crystalline and amorphous states. In order to assess the applicability of the EA-based computational methodology it is necessary to perform a detailed comparison of amorphous structures created by EAs and those prepared by cooling a liquid or by pressure-induced amorphization. In our study we address this question in case of a-N at Mbar pressure and compare the structure of three amorphous forms prepared in three completely independent manners.

\section{Simulation protocols}
\label{simulations}

\subsection{\textit{Ab initio} calculations}

For all DFT calculations we used the VASP \cite{VASP-1,VASP-2,VASP-3} software package along with the PAW\cite{VASP-PAW} method. The exchange-correlation energy was described by the PBE \cite{PBE} functional. All MD simulations were performed with hard pseudopotential PAW\_PBE N$_{h}$ using cutoff 700 eV and with the $\Gamma$ point only, except for the MD in the liquid cooling protocol at high temperatures where the softer pseudopotential \textsc{PAW\_PBE N} with cutoff 520 eV was employed (decompression of this structure and subsequent heating was performed with the former settings). Both pseudopotentials have 5 valence electrons. NPT simulations were performed using the Parrinello-Rahman barostat with fictitious masses of 4000, 3000 and 2000 for systems with 512, 256 and 192 atoms, respectively \footnote{For more information about the Parrinello-Rahman barostat and
Langevin thermostat see the VASP manual at https://cms.mpi.univie.ac.at/vasp/vasp \\
/Parrinello\_Rahman\_NpT\_dynamics\_with\_Langevin\_thermostat.html
}. Temperature was controlled by a Langevin thermostat with a friction coefficient for the atomic and lattice degrees of freedom equal to 5 and 4 ps$^{-1}$, respectively (in the case of the 192-atom system the latter parameter was equal to 
2 ps$^{-1}$). 
Smearing was performed with the Gaussian scheme and a smearing parameter equal to 0.05-0.2 eV.
To calculate final enthalpies in Table \ref{tab:hp_aN-props} we chose a hard pseudopotential with a cutoff of 900 eV and with a $\Gamma$ centered k-point grid with a length parameter of 20 \footnote{For more information about k-point density option see the VASP manual https://cms.mpi.univie.ac.at/vasp/vasp  \\
/Automatic\_k\_mesh\_generation.html}. Bulk moduli were evaluated by inducing a small volume change (0.03 \%) and calculating the finite difference of pressure. For this calculation we used the hard pseudopotential with a cutoff of 900 eV and a $\Gamma$ centered k-point grid with a length parameter of 30.

%%%%%%%%%%%%%%%%%%%%%%%%%%%%%%%%%%%%%%%%%%%%%%% glass
\subsection{a-N obtained as glass}

\begin{figure}[h]
\includegraphics[width=\columnwidth]{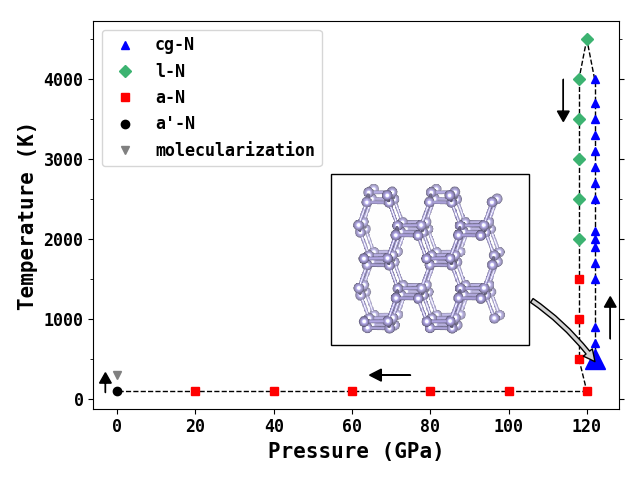}
\caption{The simulation protocol in which a-N was obtained as glass. First, the cg-N phase (blue points) was melted into a high-pressure polymeric liquid (green) at 120 GPa and a-N (red points) was obtained upon cooling down to 1500 K. This glassy amorphous form was cooled down to 100 K at 120 GPa, after which it was decompressed to 0 GPa. Finally, the decompressed form was heated in order to observe the molecularization process when energy is released (see Sec. \ref{molecularization}).}
\label{glass-protocol}
\end{figure}

The first method we used to obtain a-N  was the simulation of the standard process of glass formation by quenching of a liquid. A similar method was used in Ref.\cite{Boates-Bonev-N-2011}. We started from a 512-atom sample of the cg-N 
phase structurally relaxed at 120 GPa and gradually heated the system in $NPT$ MD simulations until the crystal melted at 4500 K.  Afterwards we cooled the system down to 1500 K where the liquid froze and an amorphous polymeric state a-N was created. The cooling rate was equal to 43 K/ps.
The simulation protocol is shown in Fig. \ref{glass-protocol}. For better comparison with other methods used to prepare a-N which employ smaller supercells we also applied a very similar liquid-cooling protocol to a smaller 256-atom system (see Table \ref{tab:hp_aN-props}).

The principal structural information is contained in N-atom coordinations, where single-coordinated N atoms are associated with molecules ($1m$) or free chain endings ($1c$), two-coordinated ($2c$) atoms with internal chain segments, and three-coordinated ($3c$) atoms act as nodes of the network (mainly cross-links for $2c$ chains). The radial limit for coordination was based on the first minimum of radial distribution functions and we chose to keep it equal to 1.8 \AA \ throughout the calculations.

\begin{figure}[h]
\begin{tabular}{cc}
\includegraphics[width=0.9\columnwidth]{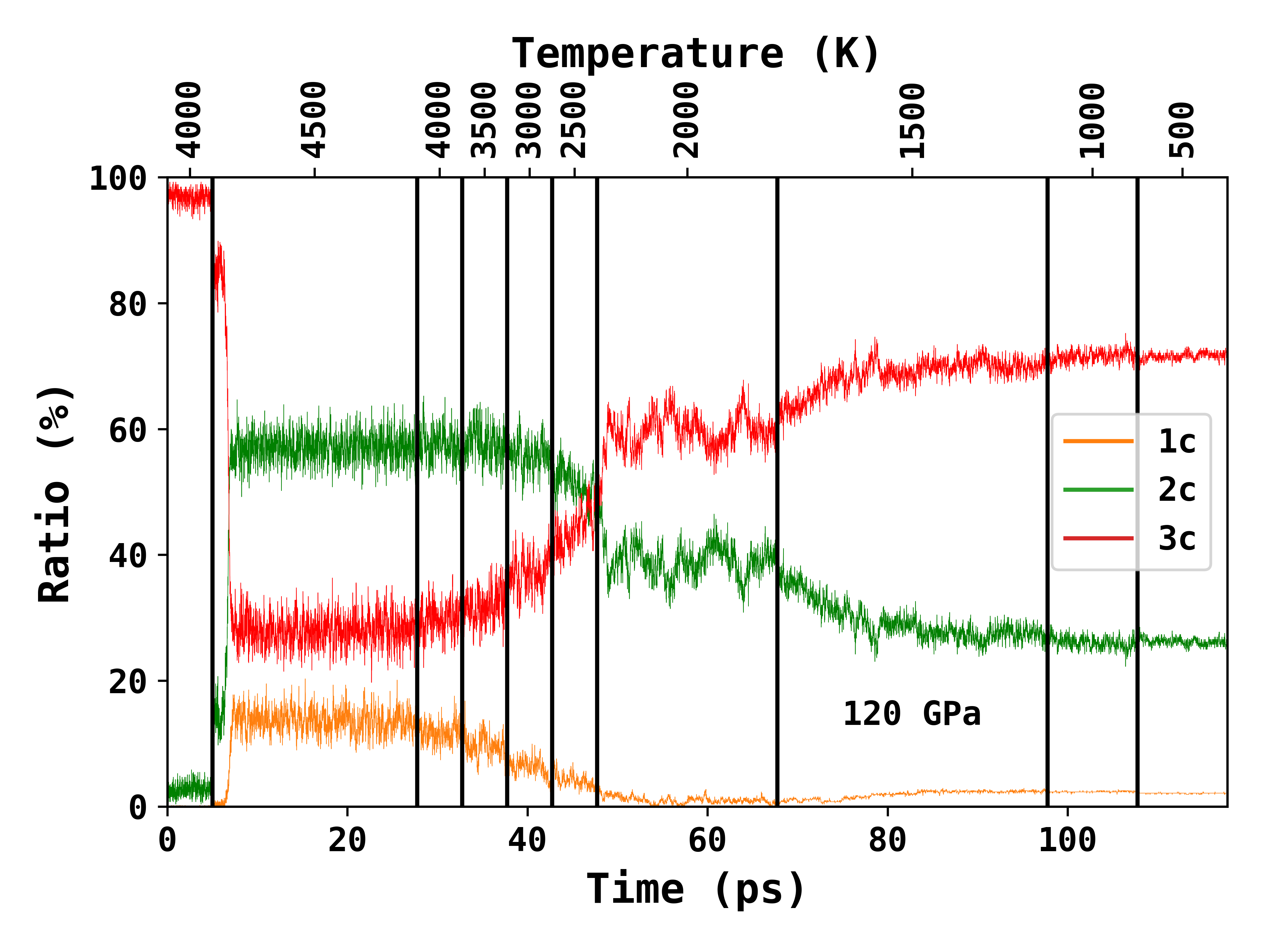}\\
\includegraphics[width=0.9\columnwidth]{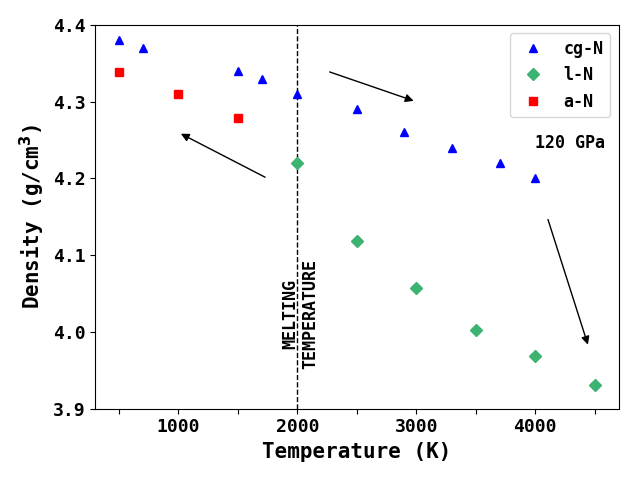}\\
\includegraphics[width=0.9\columnwidth]{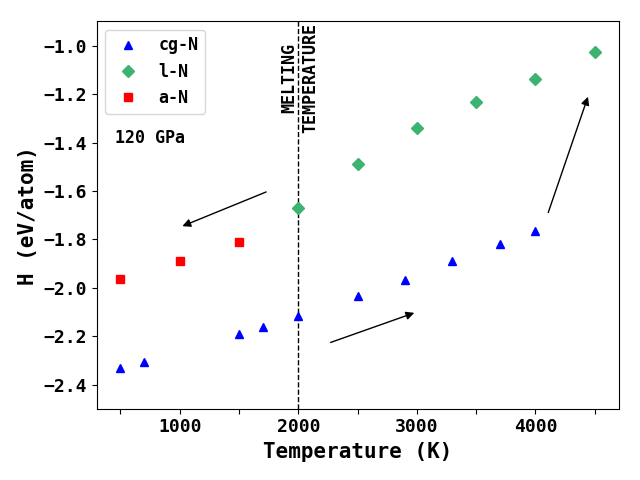}
\end{tabular}
\caption{
Evolution of nitrogen coordinations (upper panel), change in density (middle panel) and enthalpy (lower panel) during the process of glass formation: heating cg-N crystal to polymeric liquid and cooling the liquid to glass with 512 atoms.
  }
\label{glass}
\end{figure}

The corresponding evolution of N-atom coordinations and the change in density and enthalpy are shown in Fig. \ref{glass}. The liquid at high temperature above 4000 K has about 60 \% of $2c$ atoms and 30 \% of $3c$ atoms. This ratio, however, changes dramatically upon cooling and below 3000 K the number of $3c$ atoms increases substantially. Below 1500 K the number of atoms with different coordinations stabilizes with about 70\% atoms being $3c$ and 30 \% being $2c$, resulting in an average coordination of 2.7, quite close to the experimentally estimated value of 2.5 \cite{Goncharov-N-2000}. The structure of this form can be characterized by very short segments of $2c$ chains connecting the $3c$ sites which dominate the system. The density drops upon melting by about 6 \% but upon cooling down to 500 K approaches the density of cg-N and remains only about 1 \% below the latter. Concerning enthalpy, a-N below 1500 K stays about 0.4 eV/atom above the cg-N form.

%%%%%%%%%%%%%%%%%%%%%%%%%%%%%%%%%%%%%%%%%%%%%%% PIA
\subsection{a-N obtained from pressure-induced amorphization}

\begin{figure}[h]
\includegraphics[width=\columnwidth]{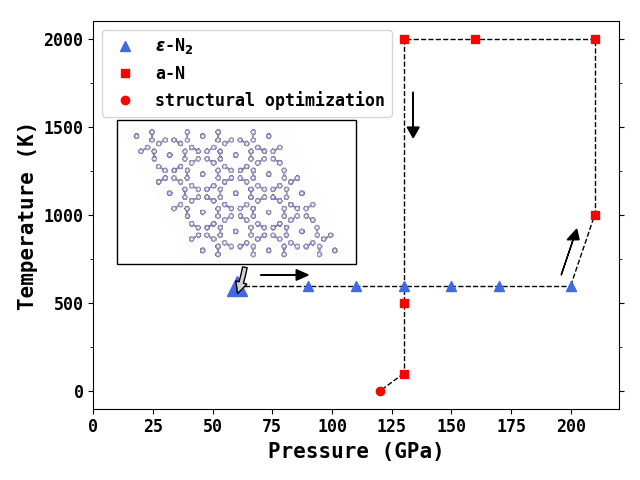}
\caption{The simulation protocol in which a-N was obtained by pressure-induced amorphization of a 192-atoms sample of the molecular $\epsilon$-N$_2$ phase shown as blue points. Polymeric a-N is shown as red points. Detailed description of the protocol is in the text.}
\label{PIA-protocol}
\end{figure}

Another way to create a-N was the simulation of the pressure-induced amorphization (PIA) process, which is the actual experimental method with which $\eta$-N was obtained\cite{Goncharov-N-2000, Eremets-N-2001, Gregoryanz-N-2001}. We started from a 192-atom sample of the molecular phase $\epsilon$-N$_2$ \cite{kotakoski2008} at 60 GPa and 600 K and gradually increased pressure to 200 GPa. Then the system was further compressed  to 210 GPa and heated to 1000 K. In this step the PIA occurred and right after the polymeric a-N was created the density dropped  by 12 \%. In order to speed up the kinetics and allow atoms to relax we heated the system to 2000 K where it was decompressed to 130 GPa during 120 ps. Afterwards the system was cooled down to 500 K and further to 100 K and finally structurally optimized at 120 GPa. This path is shown in Fig. \ref{PIA-protocol}. 

%%%%%%%%%%%%%%%%%%%%%%%%%%%%%%%%%%%%%%%%%%%%%%% evolutionary search
\subsection{a-N obtained by the evolutionary search}

We employed the code Xtalopt\cite{XtalOpt} to search for low-enthalpy structures in a supercell with 256 atoms at a pressure of 120 GPa. In order to keep simulation cell cubic-like, we imposed the following constraints: the cell vectors were chosen from the interval of 10 \AA~to 16 \AA, angles between cell vectors from the interval of 80$^\circ$ to 100$^\circ$ and finally the density from the interval of 4.11 g/cm$^3$ to 4.58 g/cm$^3$. Since we are dealing with an element we have excluded atom-exchange type operators (permustrain). We also observed that a structure-mixing type operator (crossover) at later stages of the search hardly generated better structures than random search and therefore we mainly used the stripple operator consisting of strain and wave-like displacement of atoms. We generated a total of 4600 structures within 39 generations and no crystalline structure was found (not even with local crystalline order).

\section{Results}
\label{results}
%%%%%%%%%%%%%%%%%%%%%%%%%%%%%%%%%%%%%%%%%%%%%%% comparison
\subsection{Comparison and properties of amorphous structures at high pressure}

\begin{figure}[h]
\includegraphics[width=\columnwidth]{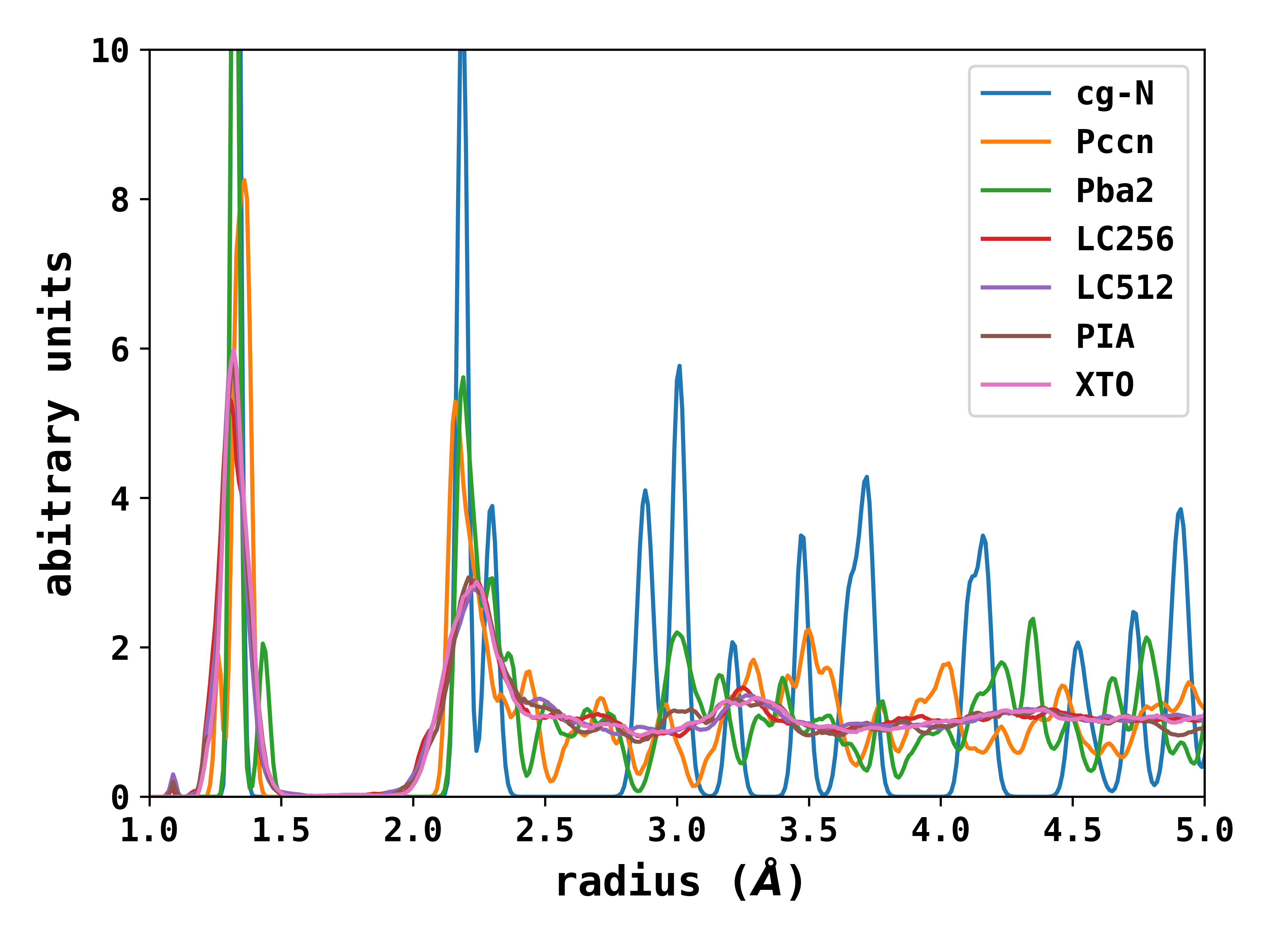}
\includegraphics[width=\columnwidth]{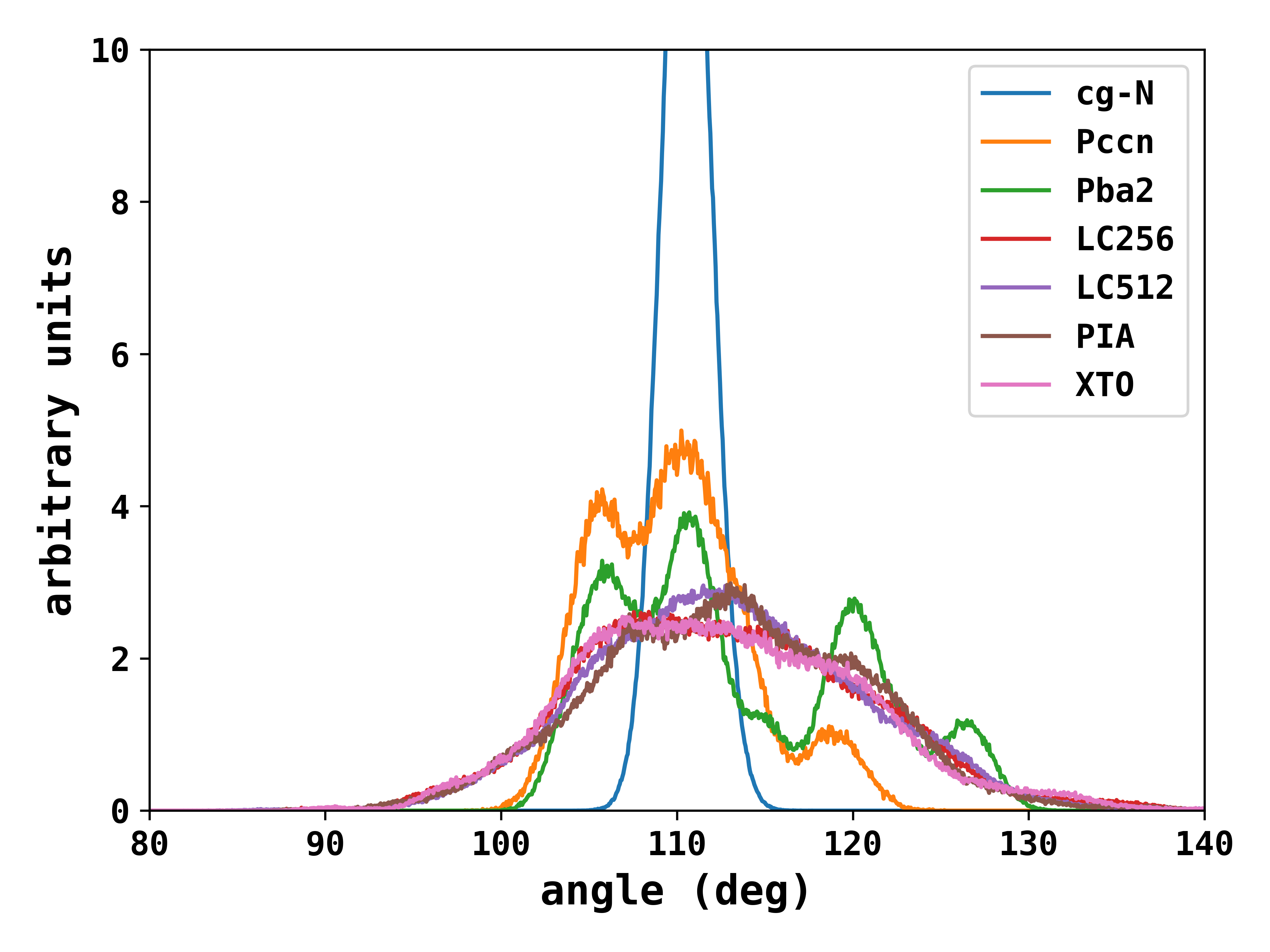}
\caption{Radial (top) and angular (bottom) distribution functions of a-N prepared by liquid cooling with 256-atom cell (LC256) and 512-atom cell (LC512), pressure-induced amorphization (PIA) and evolutionary search (XTO). For comparison, cg-N and two other metastable polymeric crystalline structures are included.}
\label{fig:R-ADF-comparison}
\end{figure}

It is interesting to compare the samples of a-N prepared independently in three entirely different ways. In Table \ref{tab:hp_aN-props} one can see that the enthalpies, densities, and the number of atoms with different coordinations and lengths of various kinds of chains are rather similar which is reassuring. We note that each protocol was applied only once and more detailed comparison would require averaging over a larger number of samples prepared by each protocol. In Fig.\ref{fig:R-ADF-comparison} (top) we see that the radial distribution functions (RDFs) in all four samples are also very similar. For comparison we included in the figure also the RDFs of cg-N and two other theoretically proposed metastable polymeric crystalline structures \textit{Pccn} \cite{Adeleke2017} and \textit{Pba2} \cite{Ma2009} calculated from MD runs at a temperature of 100 K. We can see that polymeric a-N retains the positions of the first two major RDF peaks of cg-N; however, the position of the small third peak at 3.3 \AA \,in a-N is different from that of larger peaks in the cg-N crystal which is caused by a different geometry of the rings. Beyond that distance, RDF of a-N is essentially structureless, apart from a small and broad peak around 4.3 \AA. A more detailed discussion of RDF can be found in Supp. Mat. (Fig.1).

\begin{table}[h!]
\begin{tabular}{ |c||c|c|c|c| }
\hline
                                       & \textbf{LC512} & \textbf{LC256}  & \textbf{PIA} & \textbf{XTO} \\ \hline\hline
  \textbf{enthalpy H/atom [eV]}        &       -2.067   & -2.069 & -2.097 & -2.067 \\ \hline %<- Riadok OK
  \textbf{density [g/cm$^3$]}          &        4.380   & 4.358  & 4.433  & 4.420  \\ \hline %<- Riadok OK
  \textbf{bulk modulus [GPa]}          &        241.1   & 235.1  & 249.9  & 258.8  \\ \hline %<- Riadok OK
  \textbf{ratio of $1m$ atoms [\%]}    &         1.95   &  0.00  &  0.00  &  0.00  \\ \hline %<- Riadok OK
  \textbf{ratio of $1c$ atoms [\%]}    &         0.19   &  0.78  &  1.04  &  0.00  \\ \hline %<- Riadok OK
  \textbf{ratio of $2c$ atoms [\%]}    &        25.18   & 27.20  & 16.88  & 15.33  \\ \hline %<- Riadok OK
  \textbf{ratio of $3c$ atoms [\%]}    &        72.67   & 71.92  & 82.08  & 84.67  \\ \hline %<- Riadok OK
  \textbf{aver. coord. number}         &         2.71   &  2.71  &  2.81  &  2.85  \\ \hline %<- Riadok OK
  \textbf{aver. chain length $3c\times3c$}    &         1.48   &  1.57  &  1.57   &  1.38  \\ \hline %<- Riadok OK
  \textbf{aver. chain length $1c\times3c$}    &         0.98   & absent & absent & absent \\ \hline %<- Riadok OK
  \textbf{aver. chain length $1c\times1c$}    &        absent  & absent & absent & absent \\ \hline %<- Riadok OK
\end{tabular}
\caption{High-pressure properties of amorphous structures prepared in different ways. \textit{LC512} denotes the structure prepared by liquid cooling with 512 atoms in the simulation cell, \textit{LC256} the structure prepared by a similar protocol with 256 atoms in the simulation cell, \textit{PIA} stands for pressure-induced amorphization and \textit{XTO} refers to the structure prepared by evolutionary algorithm. Chain type $3c \times 3c$ means that two $3c$ atoms are connected to each other via chain of $2c$ atoms, similarly $1c \times 3c$ represents chain with one $1c$ and one $3c$ atom at its ends and finally $1c \times 1c$ represents open chain with only $1c$ atoms at its ends. We note that we calculated the length of chain by counting only $2c$ atoms while $3c$ or $1c$ atoms at the ends were excluded. All statistics are at 120 GPa and 100 K, except for the bulk moduli and enthalpies which were calculated at 0 K.}
\label{tab:hp_aN-props}
\end{table}

\begin{figure}[h]
\includegraphics[width=\columnwidth]{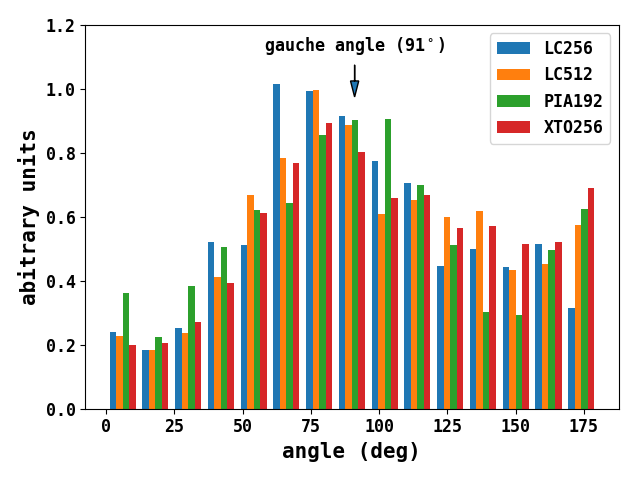}
\caption{Dihedral angles (lp-N-N-lp) for 3$c$ atoms connected by single bond in a-N prepared by liquid cooling (LC), pressure-induced amorphization (PIA) and evolutionary search (XTO).}
\label{Dihedrals-comparison}
\end{figure}

\begin{figure}[h]
\includegraphics[width=\columnwidth]{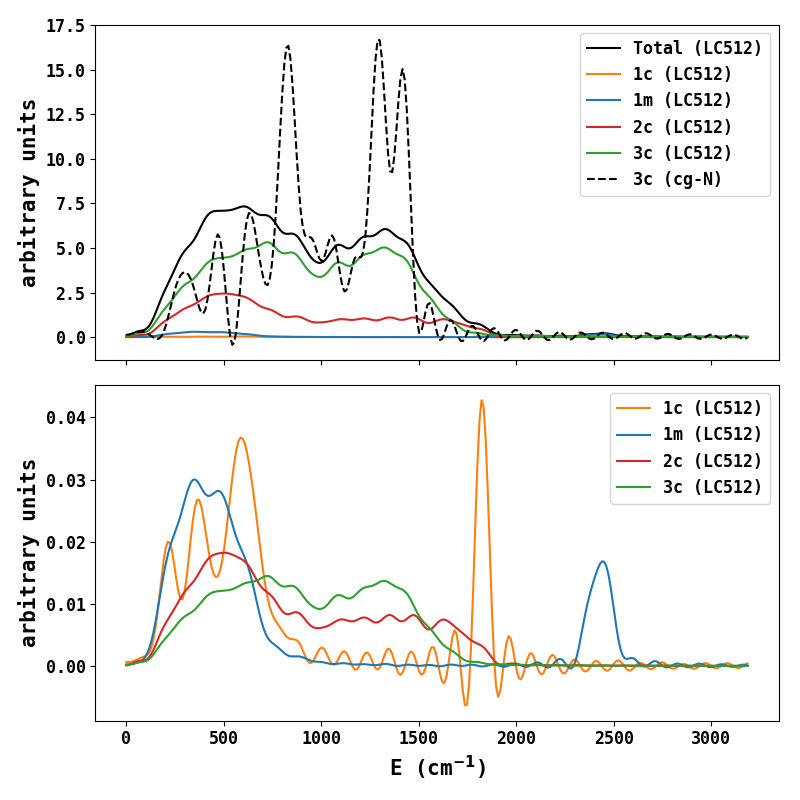}
\caption{Vibrational density of states of a-N from liquid cooling and of the polymeric crystalline cg-N phase at 120 GPa projected on atoms with different coordination (upper panel). All curves are normalized by the respective number of degrees of freedom. The lower panel shows the individual contribution of each kind of atom (see text).}
\label{vDOS-comparison}
\end{figure}

In Fig.\ref{fig:R-ADF-comparison} (bottom) we see that the angular distribution function (ADF) of all three versions of a-N is very similar as well. In particular the glassy a-N and the a-N from evolutionary search are very close while the angular distribution of a-N prepared by PIA is slightly shifted to higher angles. In all three versions the distribution spans the region from 92$^{\circ}$ to 130$^{\circ}$ and can be regarded as broadened version of the distribution in cg-N and \textit{Pccn} phases where the bond angle is close to the ideal tetrahedral angle of 109$^{\circ}$ (the ADF of crystalline structures were also calculated from MD at 100 K). Interestingly, the ADF of a-N is more similar to that of layered \textit{Pba2} structure than to that of cg-N.

An important quantity in polymeric nitrogen is the lp-N-N-lp (lp means lone pair) dihedral angle. In Ref.\cite{Radom1985} it was shown that the N-N single bond has minimal energy for the dihedral angle close to 90$^{\circ}$, i.e.,~in the gauche conformation, while the trans and cis conformations were shown to be energetically higher and much higher, respectively (see Fig.3 in Ref.\cite{Radom1985}). Subsequently, this result was used in Ref.\cite{Mailhiot1992} to identify the cg-N structure as one satisfying the condition of having all dihedral angles close to the ideal gauche value. In Fig. \ref{Dihedrals-comparison} it can be seen that polymeric a-N has instead a distribution of dihedral angles spanning the whole interval from 0 to 180$^{\circ}$ with a broad maximum around the gauche angle of 90$^{\circ}$. The relative population of cis, gauche, and trans states is in qualitative agreement with the energy curve calculated in Ref.\cite{Radom1985}. The broad character of the distribution of dihedral angles suggests that polymeric a-N does not represent a simple disordered version of cg-N. We also note that in the cg-N structure the bond and dihedral angles are strictly connected by a geometrical relation (Eq. (11), Ref. \cite{Mailhiot1992}) which prevents both angles from independently adopting their optimal values. This is probably the reason why the dihedral angle in cg-N is not so close to the optimal value of 90$^{\circ}$ but has instead a distinctly higher value of about 104$^{\circ}$ in the experimental structure \cite{Eremets-N-2004-1} (106.8$^{\circ}$ in the LDA calculation in Ref.\cite{Mailhiot1992}). Our results suggest that in disordered polymeric a-N, interestingly, both bond and dihedral angles can independently adopt their optimal values.

We also calculated the bulk modulus of the a-N prepared by the three protocols at 120 GPa. We employed the formula $B = -V \frac{\partial P}{\partial V}$ and numerically calculated the derivative by introducing small isotropic deformation. The results are included in Table \ref{tab:hp_aN-props}. For comparison the bulk modulus of cg-N at the same pressure of 120 GPa is 300-340 GPa \cite{Eremets-N-2004-1}. Our results show that similarly to cg-N also a-N is a very hard material.

An important information about disordered structure is provided by the vibrational density of states (VDOS). We calculated this quantity for a-N prepared by liquid cooling and for comparison also for cg-N from Fourier transform of the velocity autocorrelation function from a 10 ps MD run at temperature 100 K and pressure of 120 GPa. The results are shown in Fig. \ref{vDOS-comparison} (upper panel) where one can see that there is some similarity between the total VDOSs of cg-N and a-N. 
The latter, however, has some fraction of double bonds which are likely to be responsible for nonzero density in the region beyond 1500 cm$^{-1}$ where the VDOS of cg-N (containing only single bonds) already drops to zero. Projection on atoms with different coordination confirms that this feature originates mainly from 3c atoms which make a dominant contribution to the total VDOS. For convenience, we show in Fig. \ref{vDOS-comparison} (lower panel) also the projected VDOS normalized by the respective number of atoms with the given coordination which shows more clearly the individual contribution of each kind of atom irrespective of their number.

A specific quantity characterizing medium-range order in an amorphous system is the ring statistics. While in cg-N all rings consist of 10 atoms, in disordered a-N one may expect also other ring sizes. We employed the \uppercase{r.i.n.g.s} software \cite{rings} using the primitive rings method with a bond detection length of 1.8 {\AA} and found that a-N has a broad distribution of ring sizes roughly centered around the ring size of 10 (see Fig. 2 in the Supplemental Material).

\begin{figure}[h]
\includegraphics[width=0.9\columnwidth]{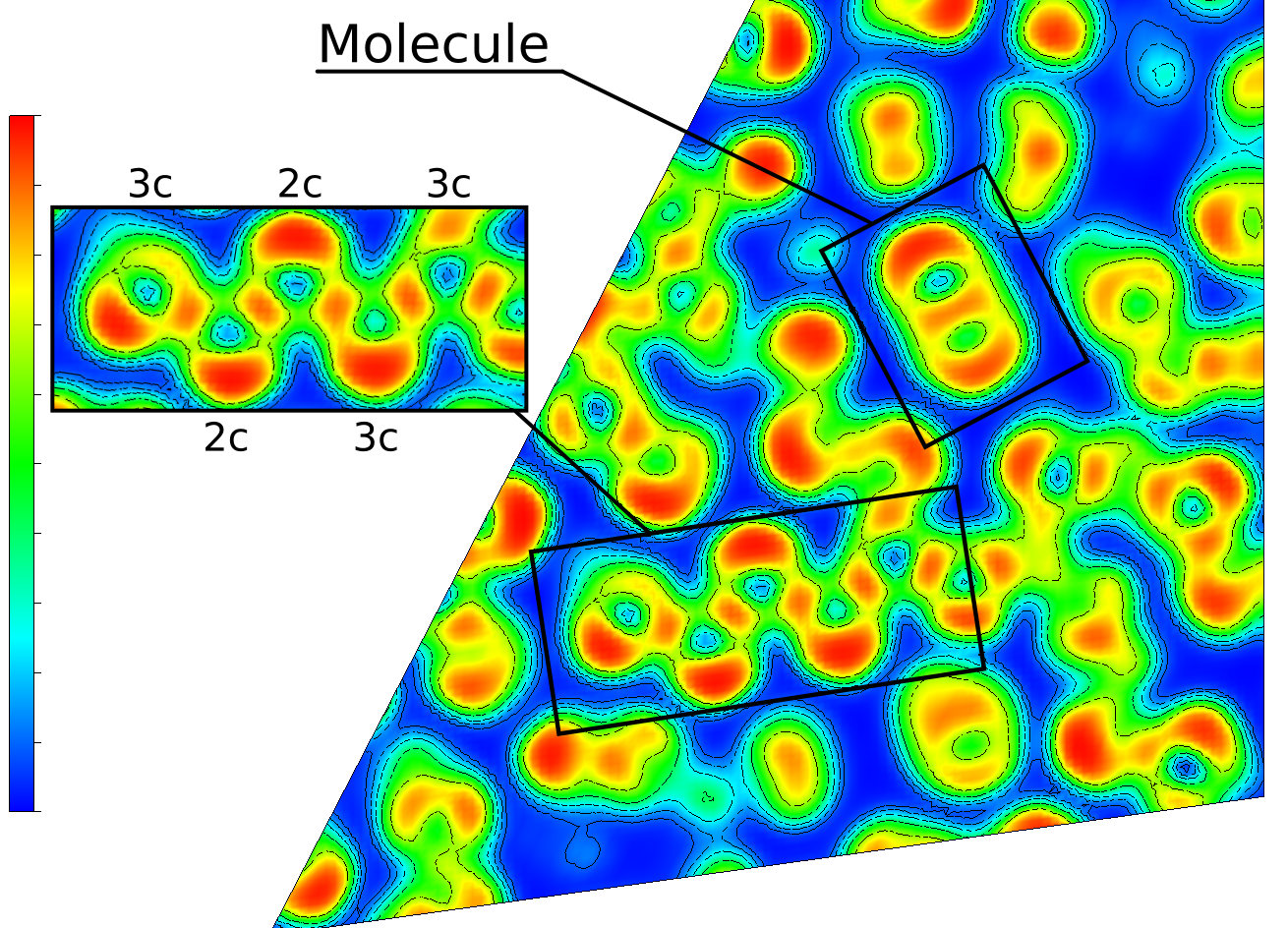}
\caption{A cut through a-N structure with 512 atoms prepared by liquid cooling at 120 GPa, where the electron localization function is shown. Visualization was made with the VESTA package\cite{momma2011}.}
\label{fig:LC:ELF}
\end{figure}

We analyzed also electronic properties and the bonding pattern of a-N at 120 GPa. We calculated the electron localization function (ELF) \cite{ELF} which is shown in Fig. \ref{fig:LC:ELF} where one can see the bonds as well as the lone pairs. In experiment \cite{Eremets-N-2001} it was found that polymeric a-N is semiconducting in the broad range of pressures from below 100 GPa up to beyond 240 GPa. We calculated the electronic density of states (e-DOS) of three a-N samples (except for the LC512 sample) prepared at 120 GPa employing the recent meta-GGA SCAN functional\cite{SCAN}. Our calculations show that the e-DOS has a semimetallic character which might also be due to some underestimating of the band gap by the SCAN functional (see Fig. 3 in the Supplemental Material).

%%%%%%%%%%%%%%%%%%%%%%%%%%%%%%%%%%%%%%%%%%%%%%% decompression
\subsection{Decompression of a-N to $p=0$ at 100 K}
\label{decompression}

\begin{figure}[h]
\includegraphics[width=\columnwidth]{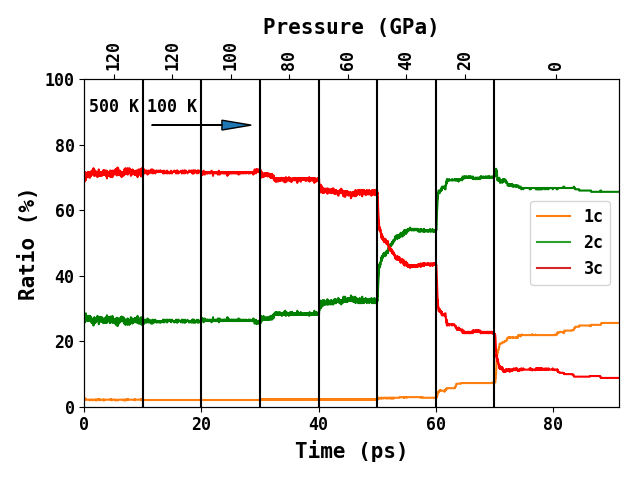}
\caption{Evolution of coordination numbers of a-N prepared by liquid cooling with 512 atom cell upon decompression from 120 GPa to $p=0$.}
\label{fig:coord-a-N-decompression}
\end{figure}

\begin{figure}[h]
\includegraphics[width=\columnwidth]{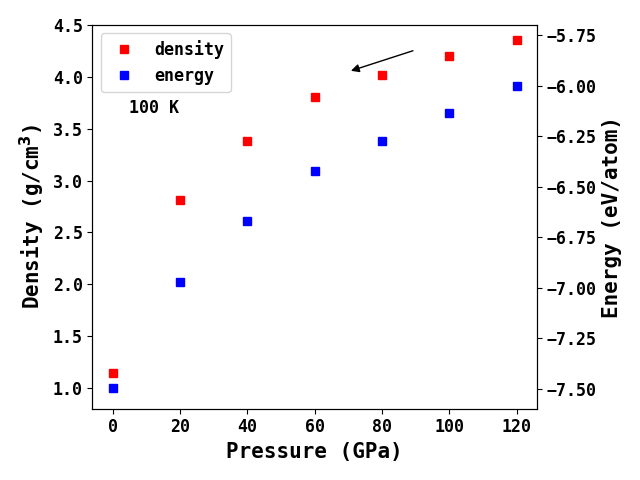}
\caption{Evolution of nitrogen density and energy during the decompression of a-N prepared by liquid cooling with 512-atom cell.}
\label{fig:density-energy-a-N-decompression}
\end{figure}

\begin{figure}[h]
\includegraphics[width=\columnwidth]{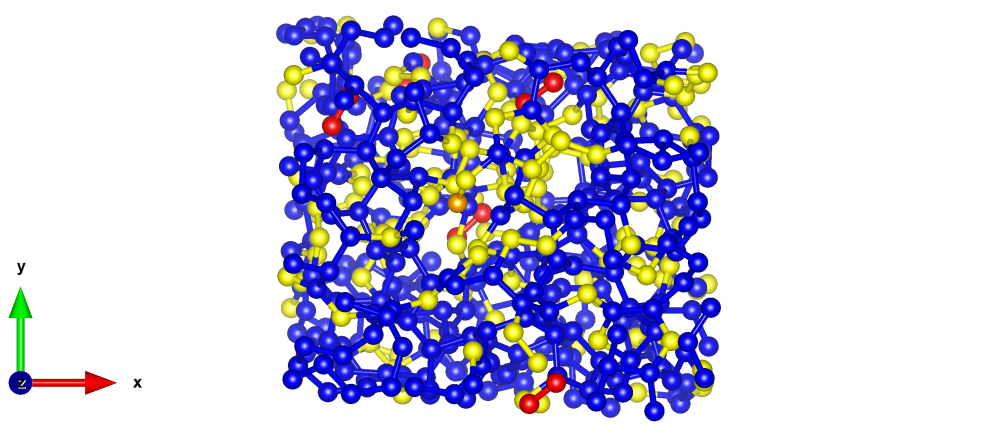}
\includegraphics[width=\columnwidth]{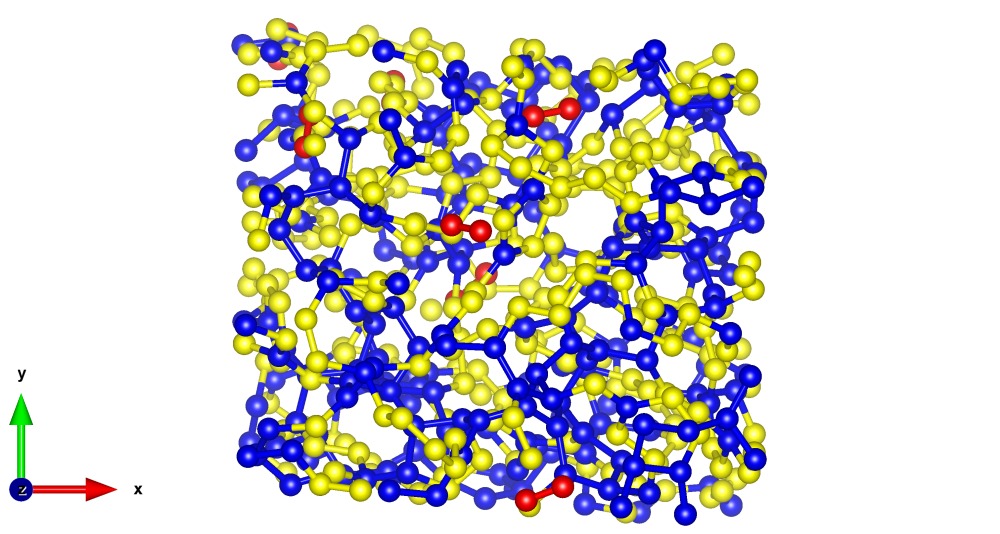}
\includegraphics[width=\columnwidth]{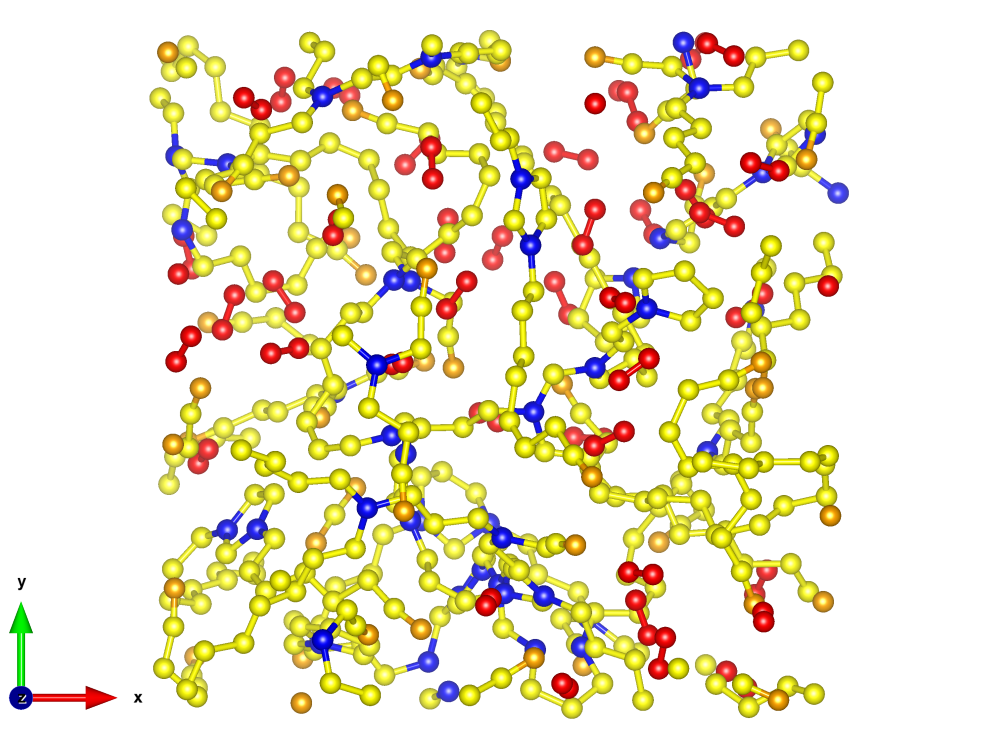}
\caption{Atomic configurations from the decompression of glassy a-N with 512 atoms at $T=100$ K at different pressures: in the top figure is the structure at 120 GPa with density 4.361 g/cm$^{3}$, the middle one represents the structure at 40 GPa with density 3.376 g/cm$^{3}$ and in the bottom figure is the structure decompressed to $p=0$ with density 1.149 g/cm$^{3}$ (a'-N). Different colors represent different coordinations of N atoms. Blue color represents the 3-coordinated atoms, yellow the 2-coordinated and orange the 1-coordinated. Red color represents the 1-coordinated atoms within a molecule. Visualization was made with the VESTA package\cite{momma2011}.}
\label{fig:a-N-coord}
\end{figure}

After characterizing the local structure of polymeric a-N at high pressure we studied its structural evolution upon low-temperature decompression, motivated by the experiment of Eremets \textit{et al.} \cite{Eremets-N-2001}, where a-N was in one case successfully decompressed to ambient pressure at temperature below 100 K. We chose to take the glassy 512-atom sample of a-N as the representative one and brought it down to $p=0$ at $T=100$ K in steps of $\Delta p = 20$ GPa (see also Fig. \ref{glass-protocol}). Evolution of nitrogen coordinations along the process is shown in Fig.\ref{fig:coord-a-N-decompression}. Throughout the decompression, we observed two rather sharp changes in coordinations upon change of pressure from 60 to 40 and from 20 to 0 GPa. We believe that the apparent sharp character of these changes is related to the fast decompression in the simulation and it is likely that if we could decompress the system more slowly and with much smaller pressure steps $\Delta p$ the evolution would be more gradual. The first change is related to the dramatic drop of the number of $3c$ atoms and increase of the number of $2c$ atoms. This correlates with the change of slope of the density and energy (Fig. \ref{fig:density-energy-a-N-decompression}) curves revealing a major structural change in the system. This structural transformation can also be seen in Fig.\ref{fig:a-N-coord} where the presence of longer chains made of $2c$ atoms is quite visible at 40 GPa. Starting at 20 GPa we also observe the onset of partial molecularization of the system (Fig.\ref{fig:coord-a-N-decompression}) that becomes even more pronounced at 0 GPa. The number of $3c$ atoms continues to decrease while the number of $1c$ atoms increases by roughly the same amount. Interestingly, during this process the number of $2c$ atoms does not change much. This, however, does not imply that $2c$ atoms do not change coordination. Instead, two transformation processes proceed at the same time and roughly at the same rate: $3c$ atoms turn into $2c$ atoms and $2c$ atoms become $1c$ atoms. As can be seen in Fig.\ref{fig:a-N-coord} the final amorphous form of nitrogen decompressed to ambient pressure (a'-N) is significantly less dense and structurally very different from the high-pressure forms.

The decompressed a'-N form contains around 20\% of $1c$ atoms, some of which represent N$_2$ molecules that are stable inside the voids of a polymeric network made up of $2c$ and $3c$ atoms. The structure of the network is based on the presence of $3c$ atoms which act as nodes connected by substantially longer (compared to a-N at 120 GPa) chain segments made of $2c$ atoms (see Fig.\ref{fig:a-N-coord}). We note that the decompressed a'-N can be regarded as a nonequilibrium amorphous structure, similarly to high-density-amorphous water ice decompressed to p=0 at liquid nitrogen temperature \cite{Mishima1984}. It is in fact a doubly metastable structure, with respect to the equilibrium amorphous structure at p=0, which is disordered molecular and also with respect to crystallization. Even though the energy difference between the fully $3c$ single-bonded cg-N and molecular N$_2$ form at $p=0$ is predicted to be as high as 1.4 eV/atom \cite{Uddin}, this value can hardly be expected in experiment since cg-N has not been successfully decompressed to ambient pressure. 
In decompressed a'-N a lower value should be expected due to the existence of a large number of $2c$ and $1c$ atoms resulting in a much smaller fraction of single bonds. We calculated the energy difference between our a'-N form and $\alpha$-N$_2$ molecular crystal at $p=0$ and found it to be 0.87 eV/atom which represents about 60\% of the ideal value of 1.4 eV/atom.

%%%%%%%%%%%%%%%%%%%%%%%%%%%%%%%%%%%%%%%%%%%%%%% molecularization

\subsection{Molecularization of a'-N after heating}
\label{molecularization}

\begin{figure}[h]
\includegraphics[width=1.0\columnwidth]{{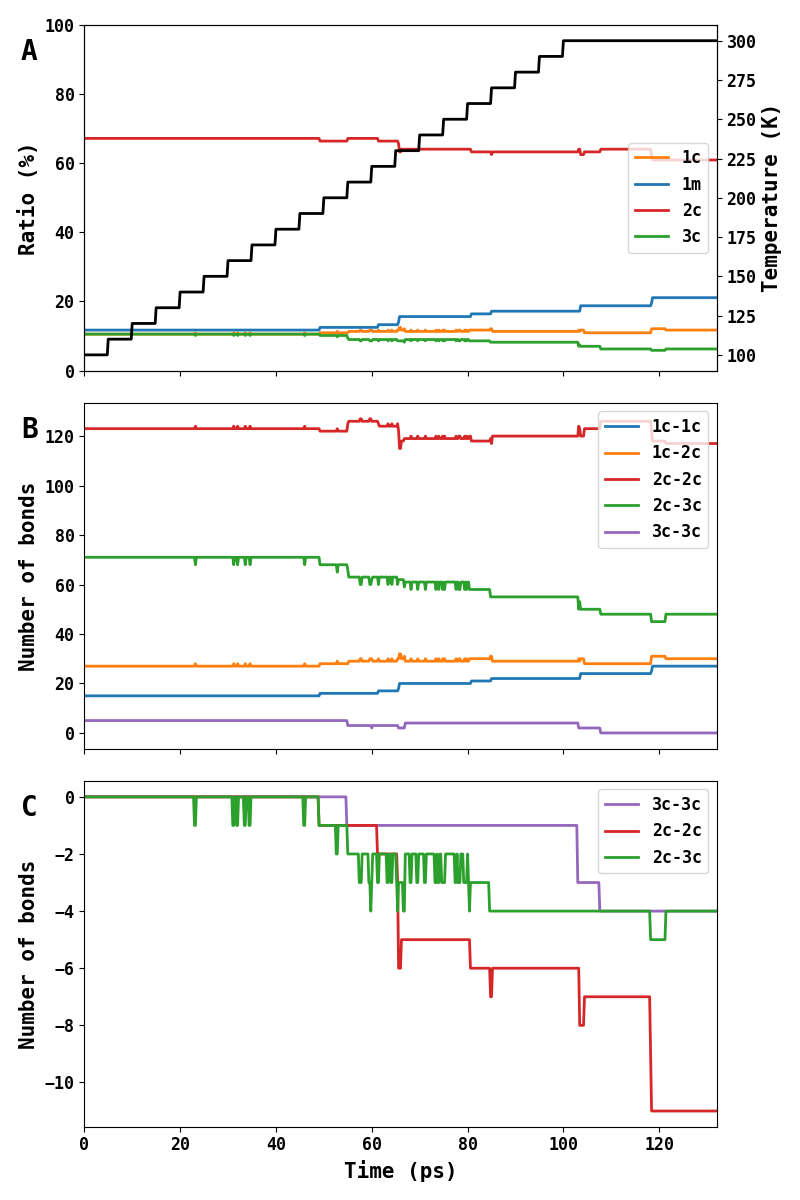}}
\caption{Evolution of atomic coordination and bonding during heating of a'-N from 100 K to 300 K at $p=0$. 
Panel (a): Temperature and fraction of $1m$, $1c$, $2c$ and $3c$ atoms ($1m$ refers to $1c$ atom forming diatomic molecule rather than end of chain). Panel (b): Evolution of total number of different bond types (1c-3c bonds are not present). Panel (c): Evolution of bond creation/breaking difference (creation and breaking of 1c-2c 
and 1c-1c bonds is not observed). 
}
\label{fig:molecularization}
\end{figure}

Since one would like to preserve at ambient pressure as much of the polymeric structure as possible it is interesting to investigate the thermal stability of the decompressed a'-N sample and the mechanism of its molecularization. To this end we decompressed the LC256 sample down to $p=40$ GPa during 60 ps in steps of 20 GPa. Further decompression to $p=0$ was in this case performed more slowly during 92 ps in steps of 2 GPa resulting in an a'-N sample. Since at low temperatures we cannot observe the molecularization process on its natural time scale which is too long for MD simulation, we chose to work at higher temperature and gradually heated the system from 100 K to 300 K during a 130 ps run. The structure in our simulation did not entirely convert into a molecular form; however, the fraction of molecules increased significantly from 12\% to 21\% (see Fig.\ref{fig:molecularization} A). Most of the atoms in the structure remained $2c$ (change from 67\% to 60\%); nevertheless the fraction of $3c$ atoms decreased from 11\% to 6\%, while that of the 1-coordinated chain ends (excluding molecules) increased only marginally from 11\% to 12\%. To analyze the mechanism closer we show in Fig.\ref{fig:molecularization} the evolution of the number of different bond types (B) and the total number of broken/created bond types (C)\footnote{We note that the creation or breaking of certain bond is not the only way how the bond type can enter or leave the statistics. For example, if a $2c$-$3c$ bond breaks it does not anymore count. However, the originally $2c$ atom becomes $1c$ atom and the originally $3c$ atom becomes $2c$ atom and therefore the number of bonds other than the broken $2c$-$3c$ is also affected.}.

First we note that no $1c$-$3c$ bonds are present (Fig.\ref{fig:molecularization} b). The breaking of $1c$-$2c$ and $1c$-$1c$ bonds is not observed since it would result in non-bonded atoms which is highly unlikely in nitrogen at these temperatures (Fig.\ref{fig:molecularization} c). The largest change is seen for the $2c$-$3c$ bonds, mainly due to $2c$-$3c$ bond breaking (Fig.\ref{fig:molecularization} c). As we can see, almost all molecule creation events (increase of number of $1c$-$1c$ bonds; Fig.\ref{fig:molecularization} b) are accompanied by $2c$-$3c$ and $2c$-$2c$ bond breaking (Fig.\ref{fig:molecularization} c).  After nearly 50 ps, when the temperature reaches about 200 K, the first bonds that start to break are $2c$-$3c$ resulting in open chains.  Slightly later also $2c$-$2c$ bonds are seen to break. This latter process appears to correlate better with the increase of the number of $1c$-$1c$ bonds and therefore can be considered as the main source for the formation of molecules. Our analysis suggests that molecules are created mainly from open chains and not by directly detaching from $3c$ atoms.  

\section{Summary and conclusions}
\label{conclusions}

We prepared polymeric a-N at high pressure of 120 GPa in three different ways, including quenching from liquid, pressure-induced amorphization and evolutionary search. The structure of all three versions of a-N was found to be quite similar, consisting of mainly 3-coordinated atoms linked by a small number of short chains. While the short-range order of a-N is similar to that of polymeric cg-N, we found significant differences in the distribution of dihedral angles which is in a-N quite broad. For this reason a-N cannot be considered as a plain disordered version of cg-N. We succeeded in decompressing a-N to ambient pressure at a temperature of 100 K, in agreement with Ref.\cite{Eremets-N-2001}. Upon decompression the structure undergoes substantial changes around 60 GPa where the fraction of $3c$ atoms and of single bonds decreases and the length of zig-zag chains (made of $2c$ atoms) connecting $3c$ atoms grows. Below 20 GPa molecules start to appear. The final a'-N structure at 100 K and $p=0$ consists of a small number of 3-coordinated atoms linked by longer polymeric chains and some number of molecules is also present. We calculated the energy stored in such decompressed amorphous phase (difference between a'-N and $\alpha$-N$_2$ molecular crystal at $p=0$) and estimate it to be about 0.87 eV/atom. This actually amounts to more than half of the ideal value calculated for the cg-N phase decompressed to $p=0$\cite{Uddin}. Finally, we analyzed the process of molecularization of the decompressed amorphous phase by heating it to room temperature and show that its main mechanism is detaching of molecules from open chains.

On the methodological side, the application of EA to search for amorphous structures may represent a promising new approach to generate amorphous materials that in principle can be applied to practically every system. It could open new possibilities for generating and studying amorphous forms of, e.g.,~poor glass formers for which it was not yet possible to avoid crystallization and experimentally prepare a disordered structure. At the same time it may allow one to study the structural evolution under pressure by preparing the amorphous structure from scratch at different pressures. This would be of particular interest in connection with the phenomenon of polyamorphism which is still incompletely understood even in important compounds such as water. In contrast to the commonly used approach to the preparation of glassy structures based on cooling a liquid, the evolutionary algorithm represents an athermal process where no physical temperature is involved and therefore it might in principle converge to some form of "ideal glass", long before reaching the crystalline state.

\begin{acknowledgments}
This work was supported by the VEGA Project No.~1/0904/15 and by the Slovak Research and Development Agency under Contract No. APVV-15-0496. Calculations were performed at the Computing Centre of the Slovak Academy of Sciences using the supercomputing infrastructure acquired in ITMS Projects No. 26230120002 and No. 26210120002 (Slovak Infrastructure for High-Performance Computing) supported by the Research and Development Operational Programme funded by the ERDF.
\end{acknowledgments}

\bibliographystyle{apsrev4-1}
\bibliography{references.bib}

\end{document}